\documentclass[11pt]{revtex4}
\usepackage{graphicx}

\begin{document}
\title{Entanglement Renormalization and Holography}
\author{Brian Swingle}
\email{bswingle@mit.edu}
\affiliation{Department of Physics, Massachusetts Institute of Technology, Cambridge, MA 02139}

\begin{abstract}
I show how recent progress in real space renormalization group methods can be used to define a generalized notion of holography inspired by holographic dualities in quantum gravity.  The generalization is based upon organizing information in a quantum state in terms of scale and defining a higher dimensional geometry from this structure.  While states with a finite correlation length typically give simple geometries, the state at a quantum critical point gives a discrete version of anti de Sitter space.  Some finite temperature quantum states include black hole-like objects.  The gross features of equal time correlation functions are also reproduced in this geometric framework.  The relationship between this framework and better understood versions of holography is discussed.
\end{abstract}

\maketitle

\section{Introduction}
Hilbert space, the mathematical representation of possible states of a quantum system, is exponentially large when the system is a macroscopic piece of matter.  The traditional theory of symmetry breaking reduces this overwhelming amount of information to three key quantities: the energy (or Hamiltonian), the symmetry of the Hamiltonian, and the pattern of symmetry breaking.  However, the existence of exotic phases of matter not characterized by broken symmetry, as in the fractional quantum hall effect \cite{fqhe}, demonstrates the need for a more general theory.  Fractional quantum hall systems are distinguished by the presence of long range entanglement in the ground state, suggesting that important information is encoded in the spatial structure of entanglement.  Here I show how such a ``pattern" of entanglement can be defined and visualized using the geometry of an emergent holographic dimension.  This picture connects two new tools in many body physics: entanglement renormalization and holographic gauge/gravity duality.

Entanglement renormalization \cite{vidal_er} is a combination real space renormalization group techniques and ideas from quantum information theory that grew out of attempts to describe quantum critical points.  The key message of entanglement renormalization is that the removal of local entanglement is essential for defining a proper real space renormalization group transformation for quantum states.  This realization has permitted a compact description of some quantum critical points \cite{vidal_qcp,italians_qcp}.  Holographic gauge/gravity duality \cite{maldecena,polyakov,witten} is the proposal that certain quantum field theories without gravity are dual to theories of quantum gravity in a curved higher dimensional ``bulk" geometry.  Holography provides a way to compute field theory observables from a completely different point of view using a small amount of information encoded geometrically.  Real space renormalization is also important in the holographic framework \cite{holo_rg0,holo_rg1,holo_rg,holo_rg2}, thus hinting at a possible connection between holography and entanglement renormalization.  We will begin with entanglement renormalization and build up to the full holographic picture.

\section{Many body entanglement}

We are interested in quantifying entanglement in many body systems.  We will use the entanglement entropy $S_A$ of a subregion $A$ as a measure of the entanglement between $A$ and the rest of the system. The entanglement entropy is defined as the von Neumann entropy $S_A = - \text{Tr}_A (\rho_A \ln{\rho_A})$ of the reduced density matrix of $A$.  The reduced density matrix $\rho_A$ is given by $\rho_A = \text{Tr}_{B} \,\rho$, where $B$ is the rest of the system and $\rho$ is the system's density matrix. If the full quantum state is pure and factorizes into a product state $|\psi\rangle = |\psi_A \rangle \otimes |\psi_B \rangle$, then $A$ is not entangled with the rest of the system and $S_A = 0$.

However, the absence of entanglement is non-generic in ground states of local Hamiltonians.  Considerable evidence suggests that in most cases the entropy satisfies a boundary law: $S_A$ is proportional to the size of the boundary $\partial A$ of $A$ \cite{arealaw1}.  This simple relationship between geometry and entanglement is violated weakly in one dimensional critical systems where the entropy scales as $(c/3) \log{L}$ with $L$ the length of the region and $c$ a universal number, the central charge.  Generic quantum states strongly violate the boundary law with $S_A$ typically proportional to the size of $A$ \cite{nature_ent}, so quantum ground states are relatively lightly entangled compared to generic quantum states.

The failure of the boundary law in one dimensional critical systems suggests that renormalization group ideas might be helpful in understanding the scaling structure.  The renormalization group organizes information in the system in terms of effective descriptions as a function of scale.  Let $z$ be the coarse grained scale at which we are studying the quantum many body system.  We would like to estimate the entanglement entropy of a region by somehow adding up contributions from entangled degrees of freedom at all relevant scales.  Conceptually, we partition the degrees of freedom by scale into groups labeled by $z$, and furthermore, we take these groups to be equally spaced in $\log{z}$.  The appropriate measure for $z$ is $d\log{z} = dz/z$ \cite{wilson_rg}.

Degrees of freedom at each scale can be entangled with the region $A$, which we take to have linear size $L$ in $d_s$ spatial dimensions.  Because of the boundary law, the contribution to the entropy of $A$ from scale $z$ should be proportional to the size of the boundary $\partial A$ in units of the coarse grained scale $z$.  This factor simply counts the degrees of freedom at scale $z$ that are local to the boundary $\partial A$.  The number of entangled degrees of freedom at scale $z$ is also proportional to the measure $dz/z$ and hence the entropy should scale as
\begin{equation}
dS \sim \frac{L^{d_s - 1}}{z^{d_s - 1}} \frac{dz}{z}.
\end{equation}
The total entanglement entropy is obtained by integrating this formula from the ultraviolet cutoff $a$ to some larger length $\xi_E$.  $\xi_E$ is the length scale beyond which there is no entanglement in the quantum state, and at a quantum critical point this length will diverge.  In one dimension the entropy also diverges with $\xi_E$, and cutting this divergence off with region size $L$ naturally gives a weak violation of the boundary law.

\section{Lattice implementation}

To obtain a more concrete formulation of entanglement as a function of scale, we will focus on a concrete lattice system where entanglement renormalization can be carried out numerically \cite{vidal_qcp}.  The quantum Ising model in one spatial dimension is a convenient model system, but the basic story applies to more general systems in higher dimensions.  The quantum Ising model has a Hamiltonian
\begin{equation}
H = -J\sum_{<ij>} \sigma^z_i \sigma^z_j - g J \sum_i \sigma^x_i,
\end{equation}
where $J$ sets the overall energy scale and $g$ is a dimensionless parameter we can tune.  The Hamiltonian consists of two competing pieces, and this competition gives rise to a quantum phase transition at $g=1$ between an ordered ferromagnetic phase and a disordered paramagnetic phase.  

Following the prescription of entanglement renormalization, one can implement a renormalization group transformation on the Ising ground state using unitary operators called disentanglers to remove local entanglement and isometries to coarse grain as shown in Figure 1.  Note that information can be lost during the coarse graining steps since the isometries typically contain projectors.  The resulting network of unitary and isometric tensors approximately encodes the ground state wavefunction using a multi-layered structure \cite{vidal_mera}.  Each layer corresponds to a different scale in the full theory, and the quantum state is effectively extended into an emergent dimension representing scale.  The network depends on $g$ because the ground state does.

\begin{figure}
\includegraphics[width = .9\textwidth]{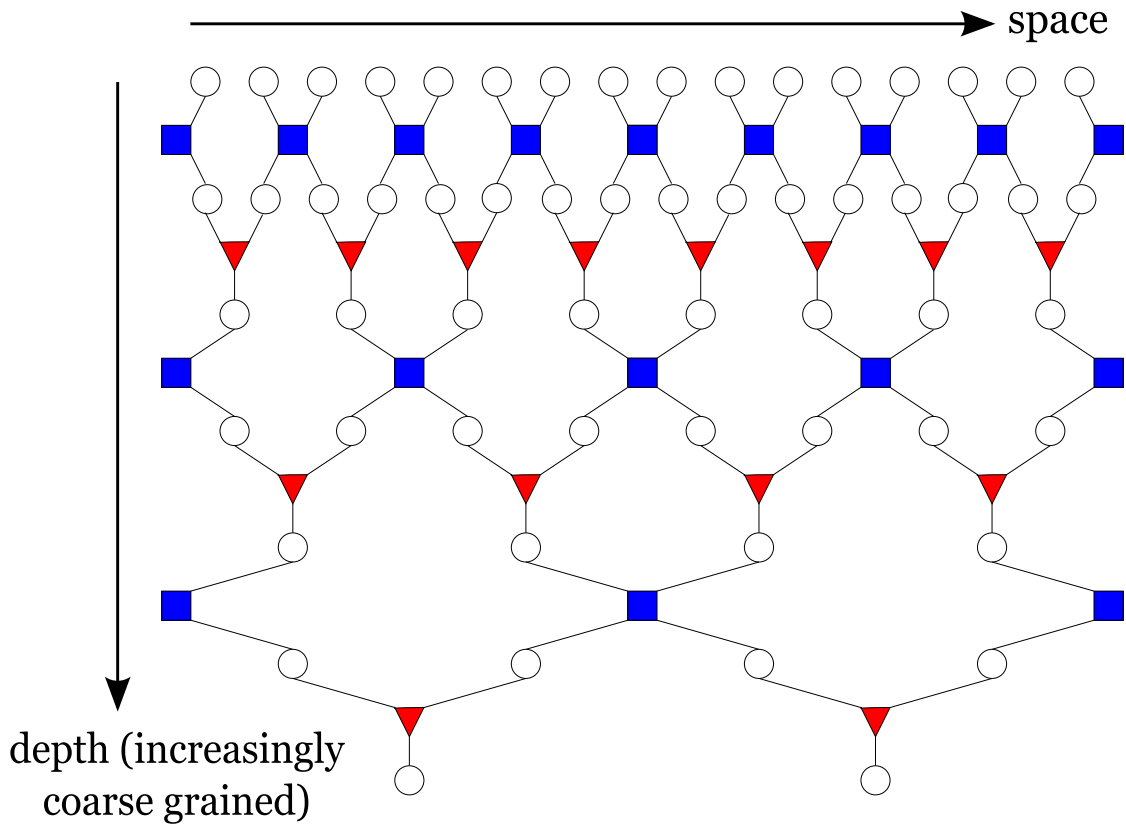}
\label{fig_1}
\caption{The tensor network structure of entanglement renormalization.  Circles are lattice sites at various coarse grained scales.  Squares with four lines are unitary disentaglers and triangles with three lines are isometric coarse graining transformations.  The network shown here represents a $2\rightarrow 1$ coarse graining scheme and has a characteristic fractal structure.  In principle, each tensor can be different, but translation and scale invariance can provide strong constraints.  This network implements a renormalization group transformation that is local in space and scale.  This transformation has the important property that it coarse grains local operators into local operators.}
\end{figure}

Inspired by holography and the connection between entropy and geometry encoded in the ordinary boundary law, we will define a geometry from the entanglement structure of the quantum state.  Imagine drawing boxes or cells around all the sites in the tensor network representing the quantum state as in Figure 2.  We view these cells as units filling out a higher dimensional ``bulk" geometry where the size of each cell is defined to be proportional to the entanglement entropy $S_{\text{site}}$ of the site in the cell.  The connectivity of the geometry is determined by the wiring of the quantum circuit represented by the tensor network in Figure 2.  The geometry ends whenever the coarse grained state completely factorizes.  Now why is such a definition useful from the point of view of entanglement?

\begin{figure}
\includegraphics[width = .9\textwidth]{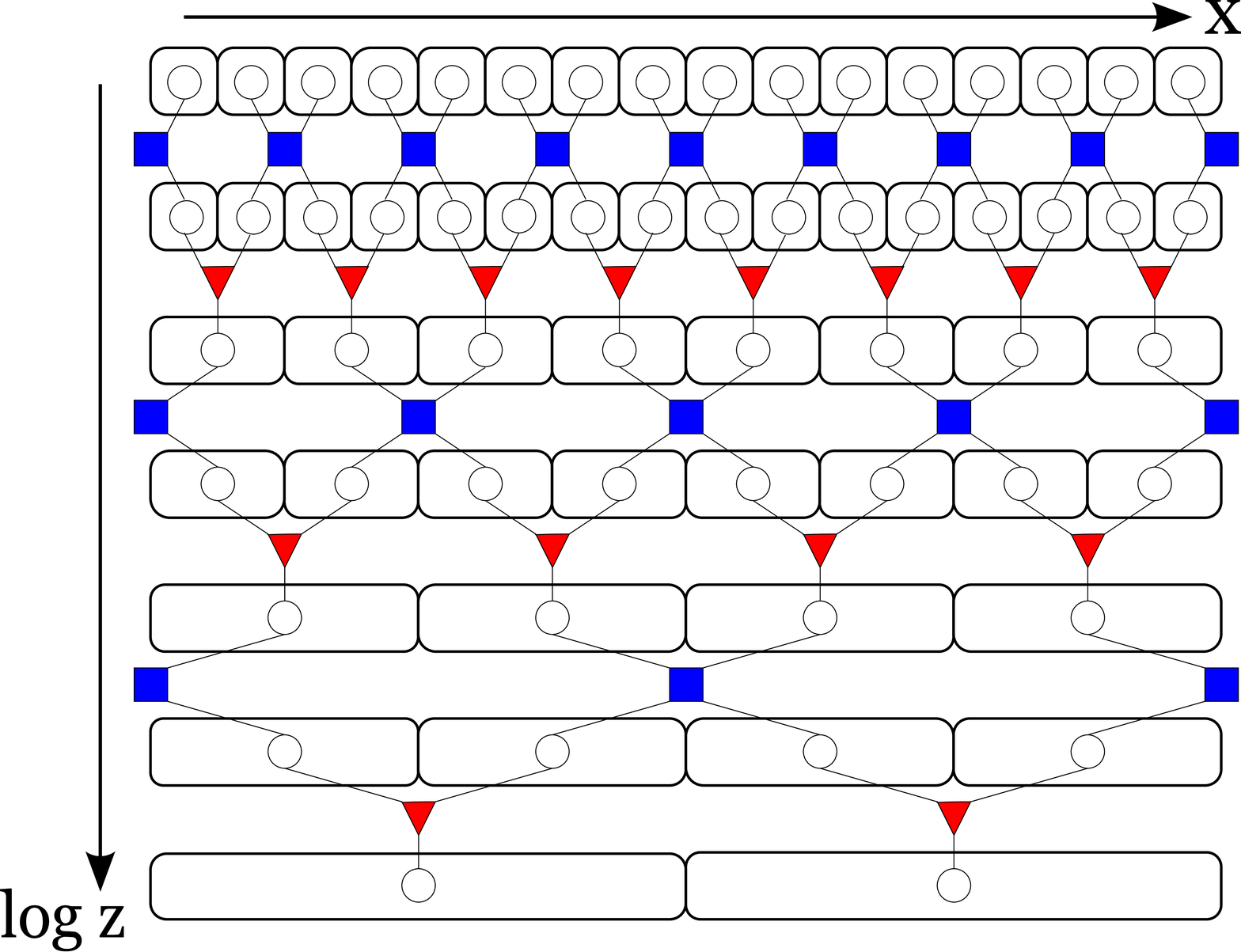}
\label{fig_2}
\caption{Curved boxes represent primitive ``cells" of the higher dimensional bulk geometry.}
\end{figure}

To compute the entropy of a block of sites in the original ultraviolet lattice we must know the reduced density matrix of the block, but what determines this density matrix?  The causal cone \cite{vidal_mera} of a block of sites in the ultraviolet is defined as the set of sites, disentanglers, and isometries that can affect the chosen block.  The causal cone should not to be confused with ordinary causality in time.  For the causal cone of a large block, the number of sites in a given layer shrinks exponentially, as in Figure 3, as we coarse grain.  Note that for a small block, say two sites, the causal cone will actually grow slightly.

We start with the density matrix for a small number of sites deep in the causal cone of the block.  The goal is to reach the ultraviolet by following the renormalization group flow backwards.  This is possible because we have recorded the entire renormalization ``history" of the state in the network, but subtleties remain because of the possible loss of information.  In practice, the truncation error may be quite small with the proper use of disentanglers.  More properly, the tensor network defines a large variational class of states for which the entanglement entropy can be computed by reversing the flow \cite{vidal_mera}.

\begin{figure}
\includegraphics[width = .9\textwidth]{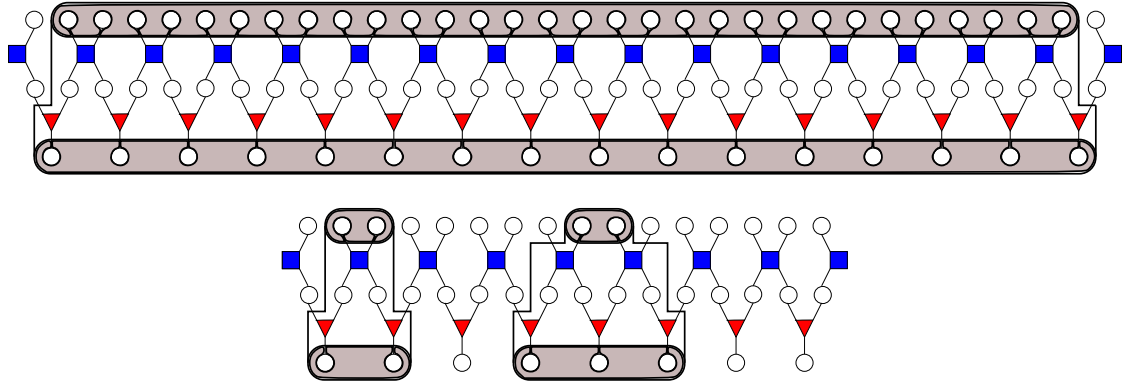}
\label{fig_3}
\caption{Causal cones of different blocks after a single layer of disentanglers and isometries. The causal cone of a large block decreases by roughly a factor of two in width (30 sites to 16 sites) after one coarse graining.  On the other hand, causal cones for small blocks may grow slightly.  Once the causal cone of a large block has reached a width of order one, it stops shrinking.}
\end{figure}

Beginning with the density matrix for a small number of sites deep in the causal cone, we reverse the isometries and disentanglers to produce the density matrix of a larger number of sites at a less coarse grained scale.  Any site at this new scale which is not in the causal cone of the block of interest can immediately be traced out as shown in Figure 4.  Tracing out a site can increase the entropy of the remainder, but the increase is no more than the entropy of the traced out site.  This procedure is repeated until the ultraviolet is reached.  Looking at the whole process, sites that are traced out occur on the outside boundary of the causal cone and form a kind of curve in the bulk geometry.  The length of this curve is by definition the sum of the entropies of all the traced out sites.  Thus the length of a curve in the bulk, including a deep ``cap" coming from the sites we began with, provides an upper bound for the entropy of a block in the ultraviolet.

\begin{figure}
\includegraphics[width = .9\textwidth]{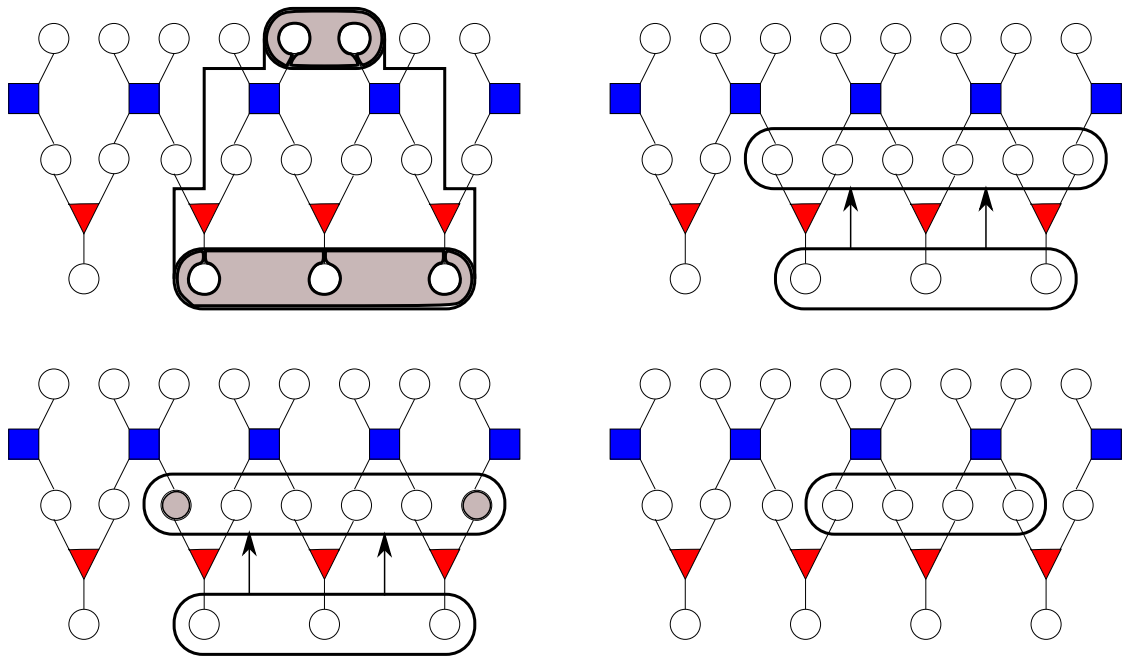}
\label{fig_4}
\caption{Upper left: a piece of the causal cone of a small block.  Upper right: reversing the flow to proceed from three sites to six sites.  Lower left: shaded sites are outside the causal cone of the two site block and can be traced out.  Lower right: four sites remain and we can now apply the next layer of disentanglers to reach the two site block of interest.}
\end{figure}

This entropy calculation is a complicated process which depends on the details of the scheme, but we can extract at least two general lessons.  First, the intuitive picture of distinct entropy contributions from each scale is realized concretely in the lattice framework.  Second, subadditivity of the entropy permits us to give an upper bound for the entropy of a block of sites in the ultraviolet.  From the cell model given above we can say that the entropy of a block of sites in the ultraviolet lattice is bounded by the length of a curve in a higher dimensional discrete geometry.  This curve is precisely a holographic screen that hides information \cite{world_holo,jac_grav}.  It is natural to view the boundary of the causal cone as a minimal curve since it represents the minimal number of sites that must be traced out, and this agrees with the geometrical definition.  As an aside, when the number of local degrees of freedom is large, similar to a thermodynamic limit, subadditivity is expected to be replaced by approximate additivity.

\section{Geometry from entanglement}

What geometries do these definitions give for the Ising model?  In the large $g$ limit, the Ising Hamiltonian is dominated by the transverse field, and the ground state is a product state.  Each site is in a pure state and we find no geometry.  Away from large $g$, the system possesses an energy gap to excited states and a finite correlation length.  The size of cells is initially non-zero due to the presence of entanglement.  However, as the quantum state is coarse grained using entanglement renormalization the correlation length should grow smaller.  Similarly, after a finite number of coarse graining steps the entanglement will be completely removed.  At this ``factorization" scale, which is the analog of $\xi_E$, the coarse grained quantum state factorizes, and we can conclude that the geometry ends.  The entanglement entropy of a block in the ultraviolet lattice receives contributions from a finite range of scales corresponding to a minimal curve hanging down from the cutoff scale to the factorization scale.  Note also that the length of this minimal curve isn't affected by the end of the geometry since the spatial length is zero there.

The geometrical picture becomes more interesting when the quantum Ising model is at its quantum critical point $g=1$.  Scale invariance forces each coarse grained layer to be identical, and the geometry continues forever.  Coarse grained sites are equivalent and have the same size.  It has been verified numerically that each coarse grained layer in the network gives an equivalent contribution to the entropy of a block, which means that the entropy is actually proportional to the length of a minimal curve \cite{vidal_qcp}.  Because of the fractal nature of the network and the equivalence of sites, the distance between points also shrinks after each coarse graining.  Entanglement renormalization is crucial for this result.  Without it we would have to keep many more states when coarse graining or otherwise settle for a poor approximation to the quantum state.

The discrete geometry that appears at the critical point is nothing but a discrete version of anti de Sitter space (AdS).  The smooth version of two dimensional anti de Sitter space has the metric
\begin{equation}
ds^2 = R^2\left(\frac{dz^2 + dx^2}{z^2}\right) = R^2 \left( dw^2 + \exp{(-2w)}\, dx^2 \right)
\end{equation}
where $R$ is some constant and $w = \log{z}$. The analog of $\log{z}$ in the lattice setup is simply the layer number or depth which counts how many coarse grainings have been performed.  The second form of the metric makes explicit the change in proper length in the spatial direction as a function of depth in the tensor network.  The appearance of AdS is perhaps not surprising from the point of view of holography, but it is gratifying to see it emerge from the definitions.

The entropy of a block in the ultraviolet is indeed bounded by the length of a minimal curve in discrete AdS because this curve counts the minimal number of entangled sites over all scales that must be traced out.  For example, we could also bound the entropy of the block by the sum of the entropies of the sites in the block, or in other words, by the length of a curve which doesn't dip into the bulk.  The length of this curve is of order $n\, S_{site} $ where $n$ is the number of sites in the block, but the length of the minimal curve is of order $\log{n} \, S_{site} $ giving a significantly better upper bound.  These bounds are similar to holographic bounds in gauge/gravity duality coming from counting degrees of freedom \cite{suss_holo}. It is also interesting to note that the minimal curve in an optimized network seems to directly control the entropy, rather than providing just an upper bound, as was confirmed in the critical case.

\section{Finite temperature and correlation functions}

Extending the picture above to finite temperature requires a shift in thinking due to the presence of classical correlations in addition to quantum entanglement at finite temperature.  The entanglement entropy now has an extensive component due to thermal effects.  However, the mutual information $I(A,B) = S_A + S_B - S_{A+B}$ between two regions $A$ and $B$ subtracts out this extensive piece and obeys a boundary law at finite temperature \cite{minfo}.  The appropriate generalization of entanglement renormalization, though it may not be as numerically tractable, is still be useful in removing local entanglement and correlations.  The coarse grained Hilbert space will typically grow if we insist on keeping all eigenvalues of the reduced density matrix of a block up to some fixed cutoff.  

Despite some peculiarities associated with hydrodynamics in quantum critical systems in one spatial dimension, we continue to focus on the quantum Ising model in one spatial dimension.  At the quantum critical point we initially find a region of discrete AdS geometry corresponding to energy scales much greater than the temperature.  However, the temperature grows as we renormalize since it represents a relevant perturbation to the quantum critical point.  Thermal effects gradually become important, and the size of coarse grained sites must begin to grow to incorporate thermalized degrees of freedom.

Eventually we reach a scale where low energy modes live.  Note that for the critical one dimensional Ising model, there is no hydrodynamic behavior for conserved currents, but the order parameter displays low energy ``quantum relaxational" dynamics \cite{subir}. What would be the ``hydrodynamic" scale is characterized by a renormalized temperature greater than the energy scale of interest but still less than the lattice scale.  If the temperature continues to grow under further renormalization then it may exceed even the lattice scale, a result familiar from the real space renormalization group of the classical one dimensional Ising model.

At this final scale the reduced density matrix of any site is proportional to the identity and the coarse grained density matrix completely factorizes.  I interpret this situation as corresponding to a black hole horizon for at least three reasons.  First, the geometry ends from the point of view of an observer ``hovering" at fixed scale.  Second, the completely mixed state is like an infinite temperature state, and the local temperature measured by a hovering observer diverges at simple black hole horizons.  Third, the final layer possesses finite entropy/size because the coarse grained sites are in mixed states.  In particular, the entropy of a large block in the ultraviolet now consists of two pieces: the usual boundary contribution plus an extensive piece due to the horizon as shown in Figure 5.

Equal time correlation functions are also interesting when viewed geometrically.  In the entanglement renormalization scheme, two operators can be correlated if the sites at which they are inserted have overlapping causal cones.  The causal cone of a single site is a ``thickened" line in the bulk geometry with a lattice scale width of a few sites.  Consider a simple gapped system.  Sites separated by less than a correlation length have overlapping causal cones, but distant sites have causal cones that end at the factorization scale before touching.  Thus distant sites cannot be correlated, and this is precisely the exponential fall off in correlations characteristic of a gapped phase.  

\begin{figure}
\includegraphics[width = .9\textwidth]{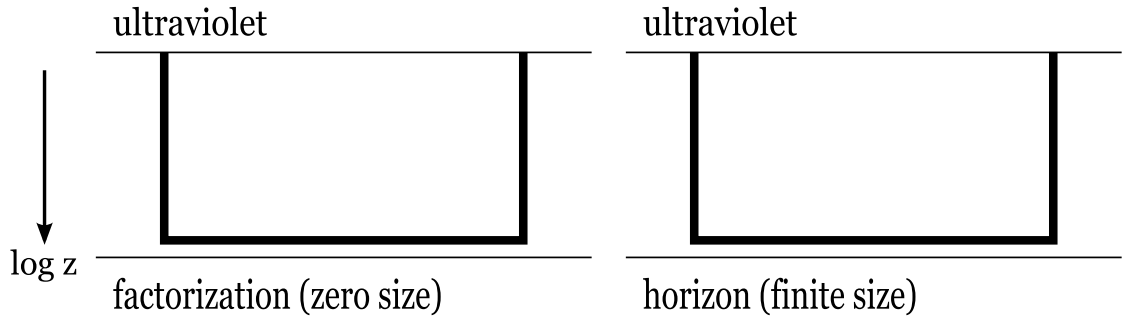}
\label{fig_5}
\caption{Sketch of minimal curves for the zero temperature gapped and finite temperature critical geometries.  The curve, defined by the boundary of the causal cone of a large block, quickly falls to the factorization scale and then runs along it.  In the gapped case, the length is zero at the factorization scale, and the entropy has only a boundary contribution which saturates for large blocks.  However, in the finite temperature case, the horizon scale has finite size and gives an extensive contribution to the entropy of a large block in the ultraviolet.}
\end{figure}

In the case of the critical geometry, the causal cones of distant sites always touch.  For conformal primary operators, which have a simple scaling behavior under renormalization, the correlation functions have an additional geometrical interpretation.  The two point function, for example, is proportional to $\exp{(-\Delta \,\ell)}$ where $\Delta$ is the operator dimension and $\ell$ is the length of a minimal curve.  This is identical to the holographic result in the conformal case.  At finite temperature, the horizon is a source of decaying correlations because the causal cones of distant sites can end at the horizon before touching.  This is nothing but thermal screening, with a screening length set by the temperature.  In each case, the structure of correlation functions is determined by the basic geometry of the extra scale dimension.

\section{Holographic duality}

The appearance of higher dimensional black holes to describe thermal states of gauge theories is precisely the content of holographic gauge/gravity duality.  The best known example of this correspondence is the duality between a certain supersymmetric quantum gauge theory in four dimensions, $\mathcal{N} = 4$ $SU(N)$ Yang-Mills theory, and a theory of quantum gravity, type IIB string theory, in a fluctuating spacetime that is asympotically five dimensional anti de Sitter space times a five sphere, AdS$_5 \times$S$^5$.  This duality is simplest from the gravity point of view when the field theory is strongly coupled and when the number of ``colors" tends to infinity, the large N limit.  In this limit quantum gravity reduces to classical gravity in a weakly curved space. 

The gauge theory in infinite volume is dual to anti de Sitter space in the Poincare patch with metric
\begin{equation}
ds^2 = R^2 \left(\frac{dz^2}{z^2} + \frac{- dt^2 + dx^2_3}{z^2}\right),
\end{equation}
where again $z$ represents something like length scale in the dual gauge theory.  Finite temperature effects in the gauge theory map to black hole mechanics in AdS, and in particular, thermal screening has an interpretation in terms of geodesics falling into the horizon.  In the classical gravity limit, the entanglement entropy of a region $A$ in the field theory is given by the area in Planck units of a minimal three dimensional surface which hangs from the two dimensional boundary of $A$ into the bulk \cite{holo_ee,holo_ee_f}.

From the point of view advocated here that entanglement defines geometry, we are free to reverse this logic and define the higher dimensional geometry in terms of entanglement entropy in the field theory.  This formulation exactly reproduces anti de Sitter space and makes the minimal surface prescription true by definition, but it may be more general since it applies to fairly generic local quantum systems.  Further evidence for the connection between entanglement renormalization and holography comes from the holographic interpretation of the extra gravitational dimension in terms of energy scale in the gauge theory.  However, this interpretation is heuristic and non-trivial because defining a real space quantum renormalization group is not straightforward.  It is tempting to regard something like entanglement renormalization as an important ingredient for making precise the real space renormalization group structure in holography.

We have seen how a geometry can be defined from entanglement, but this scheme is ultimately too simple from the point of view of gauge/gravity duality. The dual gravitational theory typically contains other fields in addition to the metric which we must also in principle define from field theory quantities.  Consistent with the interpretation of the extra dimension in terms of scale, we can attempt to define the higher dimensional bulk fields in terms of renormalized couplings in the dual field theory.  The equations of motion for the bulk fields should be taken to be the renormalization group equations for the dual field theory \cite{holo_rg}.  These couplings are typically fixed at the ultraviolet cutoff and flow under the renormalization group, but this running depends on which couplings are present and on the quantum state being renormalized \cite{holo_lor}.  One can check, for example, that relevant and irrelevant couplings in the Hamiltonian grow or decay as expected under entanglement renormalization \cite{vidal_erboson}.

Away from large N this definition is ambiguous since the bulk fields should fluctuate (the concept of a local field may even be ambiguous).  The correct prescription may be that the renormalized couplings define the expectation values of fluctuating fields which are actually quantum mechanical, a kind of quantized renormalization group.  Because entanglement renormalization permits us to address many kinds of states, it is natural to consider the renormalized couplings as dynamical variables.  This point of view is also natural in string theory.  For example, in the world sheet formulation, couplings on the string world sheet have a dual interpretation in terms of expectation values of quantum fields in the target space.  The renormalization group equations expressing conformal invariance on the worldsheet are the equations of motion for the target space fields.

One further interesting feature of gauge/gravity duality at finite temperature is that the geometry is not always equivalent to a black hole.  This understandable in our scheme because we are not guaranteed to reach a completely mixed state under a generic renormalization group flow at finite temperature.  For example, the entanglement structure of gapped states should not change dramatically due to the presence of a small temperature.  Alternatively, a conformal theory on a compact space can give at least two generic renormalization group behaviors based on whether one reaches zero spatial size or infinite temperature first. Something similar occurs in gauge/gravity duality in the form of the Hawking-Page transition for the $\mathcal{N} = 4$ theory on the space $S^3$ \cite{hp_trans}.  When a black hole does exist in the holographic geometry, the stretched horizon appears to be interpretable as the ``hydrodynamic" scale in our construction, which is naturally distinct from the ``true" (null surface) horizon.

\section{Conclusions}

We have described a framework for thinking about entanglement and correlation based on higher dimensional geometry.  We can literally construct an emergent holographic space from the entanglement properties of a large class of many body states including free bosons and fermions, quantum critical points, topological phases, frustrated quantum magnets, superconductors, and more.  The gross features of entanglement and equal time correlation functions are encoded geometrically.  This geometrical picture of entanglement is realized both in a concrete lattice setup based on entanglement renormalization and in the context of gauge/gravity duality, thus connecting these two beautiful ideas.  The theory also incorporates black hole-like objects at finite temperature that seem to share many properties with more conventional black holes in semi-classical general relativity.  What we have not done is give a detailed proposal for the gravity dual of, say, the Ising model, and if such a dual exists, it seems likely to be very complicated.  Remarkably, much of this complexity seems irrelevant for the simple geometrical ideas explored here.

There are additional features as well as many open questions.  For simplicity, we worked primarily with the quantum Ising model in one spatial dimension, but everything applies to more generic systems in higher dimensions with minor modifications.  It is also possible to include time evolution within the geometrical framework.  For example, one can determine the time component of the metric from the renormalization of the Hamiltonian.  Other interesting geometries also exist, including situations where the effective spatial dimension changes as a function of scale or where the dynamical critical exponent flows.  We are also not restricted to looking solely at ground states and thermal states, although the ``geometry" of a highly entangled state may not be anything very simple.  There are issues of non-uniqueness in the construction of the renormalization group and redundant descriptions should be interpretable in terms of bulk diffeomorphisms \cite{rg_diff}, but beyond identifying diffeomorphisms it must be possible to understand in what sense the geometry truly fluctuates away from large $N$.

We briefly comment on future directions and work in progress.  Perhaps the most pressing issue for condensed matter applications is the need for a better understanding of what lies between gauge/gravity duality as inspiration and actually having the $\mathcal{N} = 4$ plasma in the lab.  The general framework outlined here seems well suited to attacking this question.  For example, our construction should certainly apply to quantum $O(N)$ vector models, which are known to show hints of a holographic description \cite{poly_on}.  Perhaps one can make progress in understanding the dual gravitational theory.

The notion of a quantized renormalization group including gravity can be made more precise, particularly with regard to fluctuations in the bulk fields.  A certain class of topological ``string-net" phases realize exact versions of entanglement renormalization \cite{vidal_top,me_top} and provide a useful exactly solvable testing ground.  More speculatively, it would be amusing to carry out this program for lattice systems defined on some nearly translation invariant fractal, perhaps realizing a holographic dual of the epsilon expansion and giving some kind of non-integer dimensional anti de Sitter space.  Finally, from the perspective of entanglement renormalization, variational principles for the higher dimensional geometry may help simplify the search for quantum circuits to describe interesting many body states.

I would like to thank T. Lin, V. Kumar, N. Iqbal, D. Park, D. Vegh, T. Faulkner, K. Balasubramanian, T. Senthil, A. Adams, H. Liu, X.-G. Wen, L. Jones, G. Vidal, and S. Sachdev for helpful comments on this work.  A special thanks to Maissam Barkeshli and John McGreevy for many enlightening discussions on these topics.  Support for this work came from teaching undergraduates statistical mechanics and grading lots of problem sets in biophysics.

\bibliography{erhol_ref}

\end{document}